\documentclass[%
superscriptaddress,
twocolumn,
nofootinbib,
 amsmath,amssymb,nofootinbib,longbibliography,
 aps,
prd,
]{revtex4-2}

\usepackage{graphicx}
\usepackage{dcolumn}
\usepackage{bm}
\usepackage{hyperref}
\hypersetup{
    pdfnewwindow=true,    
    colorlinks=true,      
    linkcolor=blue,       
    citecolor=blue,       
    filecolor=blue,       
    urlcolor=blue         
}

\begin{document}


\title{Search for Decaying Dark Matter in the Virgo Cluster of Galaxies with HAWC}

\author{A.~Albert}
\affiliation{Physics Division, Los Alamos National Laboratory, Los Alamos, NM, USA }

\author{R.~Alfaro}
\affiliation{Instituto de F'{i}sica, Universidad Nacional Autónoma de México, Ciudad de Mexico, Mexico }

\author{J.C.~Arteaga-Velázquez}
\affiliation{Universidad Michoacana de San Nicolás de Hidalgo, Morelia, Mexico }

\author{H.A.~Ayala Solares}
\affiliation{Department of Physics, Pennsylvania State University, University Park, PA, USA }

\author{R.~Babu}
\affiliation{Department of Physics, Michigan Technological University, Houghton, MI, USA }

\author{E.~Belmont-Moreno}
\affiliation{Instituto de F'{i}sica, Universidad Nacional Autónoma de México, Ciudad de Mexico, Mexico }

\author{K.S.~Caballero-Mora}
\affiliation{Universidad Autónoma de Chiapas, Tuxtla Gutiérrez, Chiapas, México}

\author{T.~Capistrán}
\affiliation{Instituto de Astronom'{i}a, Universidad Nacional Autónoma de México, Ciudad de Mexico, Mexico }

\author{A.~Carramiñana}
\affiliation{Instituto Nacional de Astrof'{i}sica, Óptica y Electrónica, Puebla, Mexico }

\author{S.~Casanova}
\affiliation{Instytut Fizyki Jadrowej im Henryka Niewodniczanskiego Polskiej Akademii Nauk, IFJ-PAN, Krakow, Poland }

\author{J.~Cotzomi}
\affiliation{Facultad de Ciencias F'{i}sico Matemáticas, Benemérita Universidad Autónoma de Puebla, Puebla, Mexico }

\author{S.~Coutiño de León}
\affiliation{Department of Physics, University of Wisconsin-Madison, Madison, WI, USA }

\author{D.~Depaoli}
\affiliation{Max-Planck Institute for Nuclear Physics, 69117 Heidelberg, Germany}

\author{R.~Diaz Hernandez}
\affiliation{Instituto Nacional de Astrof'{i}sica, Óptica y Electrónica, Puebla, Mexico }

\author{M.A.~DuVernois}
\affiliation{Department of Physics, University of Wisconsin-Madison, Madison, WI, USA }

\author{M.~Durocher}
\affiliation{Physics Division, Los Alamos National Laboratory, Los Alamos, NM, USA }

\author{N.~Fraija}
\affiliation{Instituto de Astronom'{i}a, Universidad Nacional Autónoma de México, Ciudad de Mexico, Mexico }

\author{J.A.~García-González}
\affiliation{ITESM}

\author{M.M.~González}
\affiliation{Instituto de Astronom'{i}a, Universidad Nacional Autónoma de México, Ciudad de Mexico, Mexico }

\author{J.A.~Goodman}
\affiliation{Department of Physics, University of Maryland, College Park, MD, USA }

\author{J.P.~Harding}
\affiliation{Physics Division, Los Alamos National Laboratory, Los Alamos, NM, USA }

\author{S.~Hernández-Cadena}
\affiliation{Instituto de F'{i}sica, Universidad Nacional Autónoma de México, Ciudad de Mexico, Mexico }

\author{I.~Herzog}
\affiliation{Department of Physics and Astronomy, Michigan State University, East Lansing, MI, USA }

\author{D.~Huang}
\affiliation{Department of Physics, Michigan Technological University, Houghton, MI, USA }

\author{F.~Hueyotl-Zahuantitla}
\affiliation{Universidad Autónoma de Chiapas, Tuxtla Gutiérrez, Chiapas, México}

\author{V.~Joshi}
\affiliation{ECAP}

\author{S.~Kaufmann}
\affiliation{Universidad Politecnica de Pachuca, Pachuca, Hgo, Mexico }

\author{H.~León Vargas}
\affiliation{Instituto de F'{i}sica, Universidad Nacional Autónoma de México, Ciudad de Mexico, Mexico }

\author{J.T.~Linnemann}
\affiliation{Department of Physics and Astronomy, Michigan State University, East Lansing, MI, USA }

\author{A.L.~Longinotti}
\affiliation{Instituto de Astronom'{i}a, Universidad Nacional Autónoma de México, Ciudad de Mexico, Mexico }

\author{G.~Luis-Raya}
\affiliation{Universidad Politecnica de Pachuca, Pachuca, Hgo, Mexico }

\author{K.~Malone}
\affiliation{Space Science and Applications Group, Los Alamos National Laboratory, Los Alamos, NM, USA }

\author{J.~Martínez-Castro}
\affiliation{Centro de Investigaci'on en Computaci'on, Instituto Polit'ecnico Nacional, M'exico City, M'exico.}

\author{J.A.~Matthews}
\affiliation{Dept of Physics and Astronomy, University of New Mexico, Albuquerque, NM, USA }

\author{P.~Miranda-Romagnoli}
\affiliation{Universidad Autónoma del Estado de Hidalgo, Pachuca, Mexico }

\author{J.A.~Morales-Soto}
\affiliation{Universidad Michoacana de San Nicolás de Hidalgo, Morelia, Mexico }

\author{M.~Mostafá}
\affiliation{Department of Physics, Pennsylvania State University, University Park, PA, USA }

\author{A.~Nayerhoda}
\affiliation{Instytut Fizyki Jadrowej im Henryka Niewodniczanskiego Polskiej Akademii Nauk, IFJ-PAN, Krakow, Poland }

\author{L.~Nellen}
\affiliation{Instituto de Ciencias Nucleares, Universidad Nacional Autónoma de Mexico, Ciudad de Mexico, Mexico }

\author{M.U.~Nisa}\email{Corresponding Author: nisamehr@msu.edu}
\affiliation{Department of Physics and Astronomy, Michigan State University, East Lansing, MI, USA }

\author{R.~Noriega-Papaqui}
\affiliation{Universidad Autónoma del Estado de Hidalgo, Pachuca, Mexico }

\author{N.~Omodei}
\affiliation{Department of Physics, Stanford University: Stanford, CA 94305–4060, USA}

\author{E.G.~Pérez-Pérez}
\affiliation{Universidad Politecnica de Pachuca, Pachuca, Hgo, Mexico }

\author{C.D.~Rho}
\affiliation{SKKU}

\author{D.~Rosa-González}
\affiliation{Instituto Nacional de Astrof'{i}sica, Óptica y Electrónica, Puebla, Mexico }

\author{M.~Schneider}
\affiliation{Department of Physics, University of Maryland, College Park, MD, USA }

\author{Y.~Son}
\affiliation{UOS}

\author{R.W.~Springer}
\affiliation{Department of Physics and Astronomy, University of Utah, Salt Lake City, UT, USA }

\author{O.~Tibolla}
\affiliation{Universidad Politecnica de Pachuca, Pachuca, Hgo, Mexico }

\author{K.~Tollefson}
\affiliation{Department of Physics and Astronomy, Michigan State University, East Lansing, MI, USA }

\author{I.~Torres}
\affiliation{Instituto Nacional de Astrof'{i}sica, Óptica y Electrónica, Puebla, Mexico }

\author{R.~Torres-Escobedo}
\affiliation{SJTU}

\author{R.~Turner}
\affiliation{Department of Physics, Michigan Technological University, Houghton, MI, USA }

\author{F.~Ureña-Mena}
\affiliation{Instituto Nacional de Astrof'{i}sica, Óptica y Electrónica, Puebla, Mexico }

\author{L.~Villaseñor}
\affiliation{Facultad de Ciencias F'{i}sico Matemáticas, Benemérita Universidad Autónoma de Puebla, Puebla, Mexico }

\author{X.~Wang}
\affiliation{Department of Physics, Michigan Technological University, Houghton, MI, USA }

\author{I.J.~Watson}
\affiliation{UOS}
\author{S.~Yun-Cárcamo}
\affiliation{Department of Physics, University of Maryland, College Park, MD, USA }

\collaboration{HAWC Collaboration}

\date{\today}

\begin{abstract}
The decay or annihilation of  dark matter particles may produce a steady flux of very-high-energy gamma rays detectable above the diffuse background.  Nearby clusters of galaxies provide excellent targets to search for the signatures of particle dark matter interactions. In particular, the Virgo cluster spans several degrees across the sky and can be efficiently probed with a wide field-of-view instrument. The High Altitude Water Cherenkov (HAWC) observatory, due to its wide field of view and sensitivity to gamma rays at an energy scale of 300 GeV—100 TeV is well-suited for this search. Using 2141 days of data, we search for gamma-ray emission from the Virgo cluster, assuming well-motivated dark matter sub-structure models. Our results provide some of the strongest constraints on the decay lifetime of dark matter for masses above 10 TeV.
\end{abstract}

\keywords{Dark Matter, Cherenkov, galaxy cluster, TeV}

\maketitle


\section{\label{sec:intro}Introduction}

Galaxy clusters -- massive ($>10^{13} \rm M_\odot$) gravitationally bound conglomerates of galaxies within a few Mpc of each other -- are among the most important probes of large-scale structure in the universe. The kinematics of galaxy clusters have historically constituted an important piece of evidence for the existence of dark matter (DM) \cite{2011ARA&A..49..409A}.  The particle nature of DM, however, remains elusive. Among the predictions of various theories of physics beyond the Standard Model, Weakly Interacting Massive Particles (WIMPs) are the leading candidates for particle DM. The aforementioned particles may be indirectly detected in astrophysical surveys via the electromagnetic or neutrino signatures of WIMP self-interactions \cite{PerezdelosHeros:2020qyt}. These particles may annihilate or decay in regions of high DM density, and produce gamma rays, either directly or through the decay of intermediate standard model particles. 

As described in section \ref{subsec:dmmodel} in detail, the signal due to DM from a given region of interest is a function (among other factors) of the DM density in the region. For DM annihilation, the signal is proportional to the square of the DM density (since two particles are required for annihilation). On the other hand, for DM decay, the signal is proportional to the DM density as a single particle can decay.  In this work, we focus on DM decay for which large extended regions in the sky, such as galaxy clusters, are particularly good targets. In searches for annihilation, there is no significant advantage from large spatial extensions; point-like sources such as dwarf galaxies are better suited due to the aforementioned dependency on DM density-squared (see Refs. \cite{HAWC:2019jvm,Alfaro_2023,2018ApJ...853..154A,HAWC:2018eaa,HAWC:2018szf} for HAWC searches for annihilating dark matter from various targets). 
For WIMP masses above 1 TeV, both gamma-ray and neutrino telescopes have performed searches, scanning nearby galaxies and clusters and yielding important constraints on the decay lifetime of DM \cite{2012ApJ...761...91A,MAGIC:2018tuz,HAWC:2018eaa,IceCube:2018tkk,IceCube:2022clp,IceCube:2023ies,LHAASO:2022yxw,Maity:2021umk}. 

The Virgo cluster is the closest galaxy cluster consisting of more than 1200 known galaxies and spanning a diameter of approximately 12$^\circ$ on the sky \cite{2020A&A...635A.135K}. One of the most notable objects near the center of the cluster is the supermassive black hole in the galaxy M87. This active galactic nucleus has been known to emit TeV gamma rays via accretion \cite{Acciari_2008,Berge:2006uls,Neronov:2007vy,2020MNRAS.492.5354M}. The large apparent size of the galaxy cluster, with multiple embedded point sources within, makes it difficult to search for gamma-ray emission via Imaging Air Cherenkov Telescopes due to their small fields of view. The HAWC observatory's wide field of view makes it well-suited to probe extended objects for gamma-ray signatures of DM. In this work, we perform an analysis to search for gamma rays produced by the decay of DM through several bosonic, leptonic, and quark channels in the TeV--PeV mass range. Heavy DM particles beyond the GeV scale are well motivated in several theories; see \cite{Bauer:2020jay} and references therein for examples.  

This paper is structured as follows. In section \ref{sec:hawc}, we review the HAWC detector and data set used in this work. We describe the analysis details in section \ref{sec:analysis}, which include the spectrum of M87 and the DM models used in the search. In section \ref{sec:res}, we present our results in the form of lower limits on the decay lifetime of DM for various channels and conclude. 

\section{\label{sec:hawc}HAWC Data}
The HAWC observatory is an array of 300 water-Cherenkov detectors (WCDs), covering an area of 22,000 m$^2$, at an altitude of 4.1 km above sea level in the state of Puebla, Mexico. Recently, HAWC has been upgraded to include an additional 345 outrigger tanks, though data from the outriggers are not included in this analysis. Each WCD consists of a 4.5 m high tank filled 200,000 litres of purified water instrumented with four photomultiplier tubes. HAWC detects the secondary air showers of charged particles produced by gamma rays and cosmic rays interacting with the earth's atmosphere. The spatial and temporal distribution of charge registered by the array during an event is used to reconstruct the direction, energy and primary particle type initiating the shower. The observatory is sensitive to gamma rays of energies between 300 GeV to more than 100 TeV, achieving a hadronic background suppression of more than 99\% at the highest energies. HAWC can monitor the sky continuously with an instantaneous field of view of 2 sr, making it particularly useful for detecting extended regions of emission that subtend several degrees on the sky. More details of the detector hardware and event reconstruction algorithms can be found in Refs. \cite{historical:2023opo,2017ApJ...843...39A} .

We use 2141 days of ``Pass 5'' data, as introduced in \cite{HAWC:2022khj,Yun-Carcamo:2023Na}, collected between March 2015 and January 2021. The energies of the gamma-ray events in the data are reconstructed using a retrained neural network first described in Ref. \cite{HAWC:2019xhp}. 

\begin{figure}[h!] 
    \centering
    \includegraphics[width=1.\linewidth]{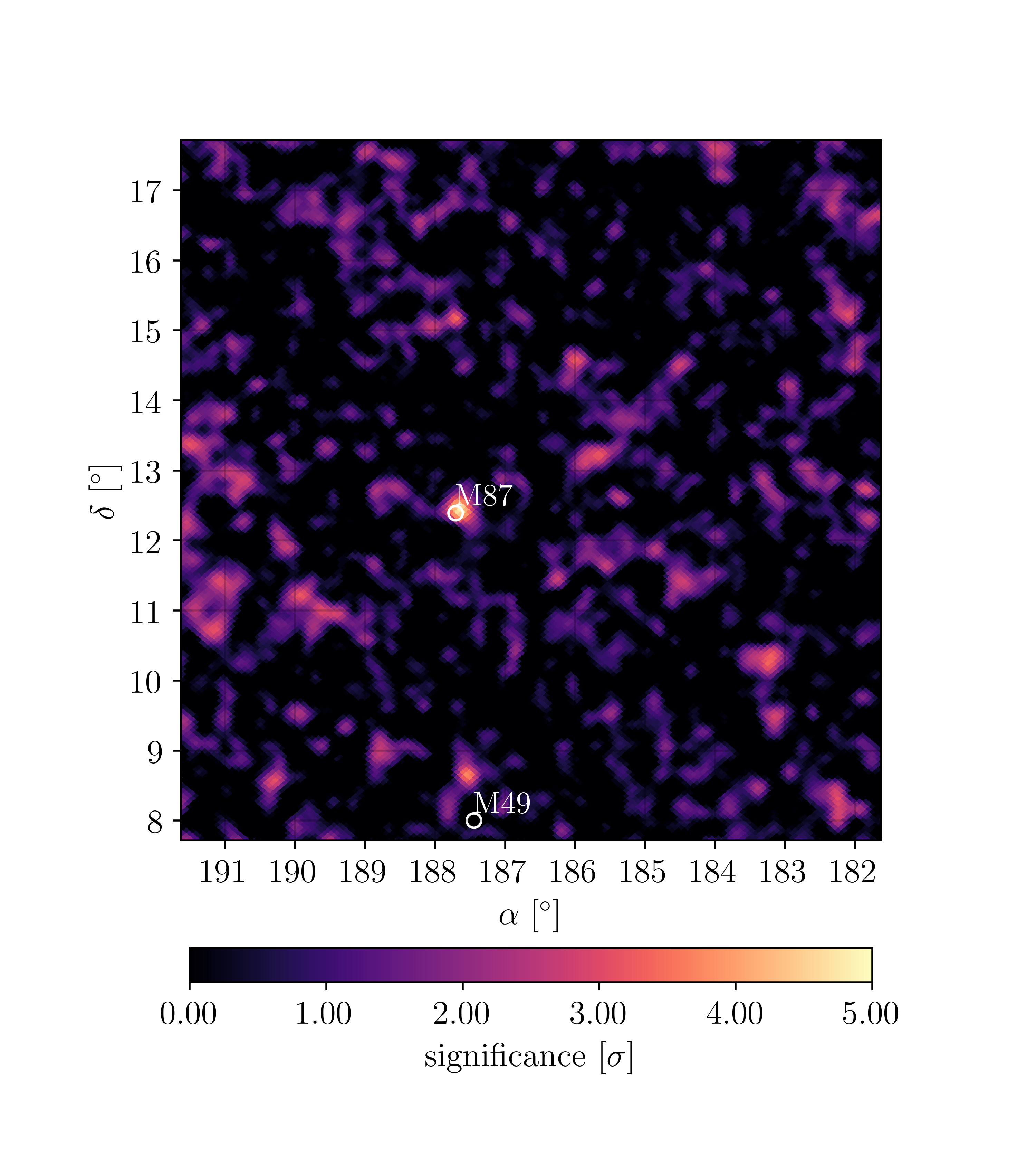}
    \caption{\label{fig:map} The point-source significance map around the Virgo cluster region using 2141 days of HAWC data with pass 5 reconstruction and the methods used in Ref. \cite{HAWC:2020hrt}. The positions of the two main sub-clusters used in this analysis are labeled as M87 and M49.}
\end{figure}

\section{\label{sec:analysis}Analysis}
In this section, we first describe the observation of the Virgo cluster region in HAWC data, including the significant point source emission from M87. We then describe the binned maximum likelihood method that is used to search for emission compatible with a DM hypothesis after accounting for the signal contamination from the direction of M87.     
\subsection{Region of Interest}
To detect point sources in the sky, we project the HAWC data on a healpix grid of NSIDE 1024, and compare the events in each pixel to the isotropic background as described in Ref. \cite{HAWC:2020hrt}. The background is estimated using the method of direct integration, which effectively convolves the all-sky rate with the detector acceptance in 2 hour periods \cite{2003ApJ...595..803A}. Figure \ref{fig:map} shows a 10$^\circ$ by $10^\circ$ map of HAWC data spanning the Virgo cluster, centered at the coordinates (RA $=186.63^\circ$, Dec $=12.72^\circ$).  The Virgo Cluster spans about 12$^\circ$ on the sky and consists of multiple galaxy groups or sub-clusters. Two of these sub-clusters, Virgo A and Virgo B,
 are centered on the galaxies M87 and M49 respectively \cite{Cote:2003er}. The other sub-cluster is centered around M86, an elliptical
galaxy \cite{1999A&A...343..420S}. The subclusters A and B  dominate the Virgo Cluster mass \cite{1999A&A...343..420S}. A $\sim 5\sigma$ excess at the position of M87 is observed in HAWC data (fig. \ref{fig:map}). No significant emission is observed at the location of M49.

\subsubsection{M87}
We determine the energy spectrum that best describes the HAWC observation of M87 by fitting the data to an attenuated power law. The spectral energy distribution $dN/dE$ can be parametrized as,
\begin{equation}
    \frac{dN}{dE} = A\left(\frac{E}{E_0}\right)^{-\gamma} \exp(-\tau_{\rm ebl})
    \label{eq:2}
\end{equation}
where $A$ is the flux normalization, $E_0$ is the reference energy fixed at 1 TeV, $\gamma$ is the spectral index. The factor $\exp(-\tau_{\rm ebl})$ is the survival probability of gamma rays as they propagate over intergalactic distances and interact with the extragalactic background light (EBL). $\tau_{\rm ebl}$ is the optical depth of the intervening medium through which the photons propagate. We use the model from Ref. \cite{Franceschini:2017iwq} to describe $\tau_{\rm ebl}$.  We note that considering the low redshift of the target, the impact of EBL attenuation is small and only significant for the gamma-ray signal above $\sim 20$ TeV. At 1 TeV, the survival probability, $\exp (-\tau_{\rm ebl})$, has a value of  $\sim 0.96$. At 100 TeV, it decreases to $\sim 0.001$. 

We use the same likelihood maximization framework as used in previous HAWC publications \cite{HAWC:2019xhp,Vianello:2015wwa}. The best-fit values of the parameters and their 1-$\sigma$ errors that describe the spectrum of  M87 are given in table \ref{tab:spec}.  The test statistic, given by the negative ratio of the best-fit likelihood, and the background-only likelihood is 35.  The flux at 1 TeV is consistent with measurements by VERITAS and H.E.S.S., within experimental uncertainties \cite{Acciari_2008,Berge:2006uls,Neronov:2007vy}.  A detailed HAWC publication on the time-dependent and multi-wavelength emission of M87 is in preparation. For this analysis, we treat it as a steady foreground source in our region of interest. 

\begin{table}[h!]
\centering
\begin{tabular}{c|c|c}
\hline
 $A \times 10^{-13}$ [$\rm TeV^{-1} \rm cm^{-2} \rm s^{-1}$] & $\gamma$ \\
\hline
$3.8 \pm 1.8$   & $2.2 \pm 0.2$ \\
\hline
\end{tabular}
\caption{The best-fit spectral normalization and index for the fit to eq. \ref{eq:2} for M87, with $E_0$ fixed at 1 TeV. The reported uncertainties are statistical.}
\label{tab:spec}
\end{table}

\begin{figure*}[t!]
\makebox[1.00\width][c]{
\begin{tabular}{@{}cc@{}}
\includegraphics[width=0.47\textwidth]{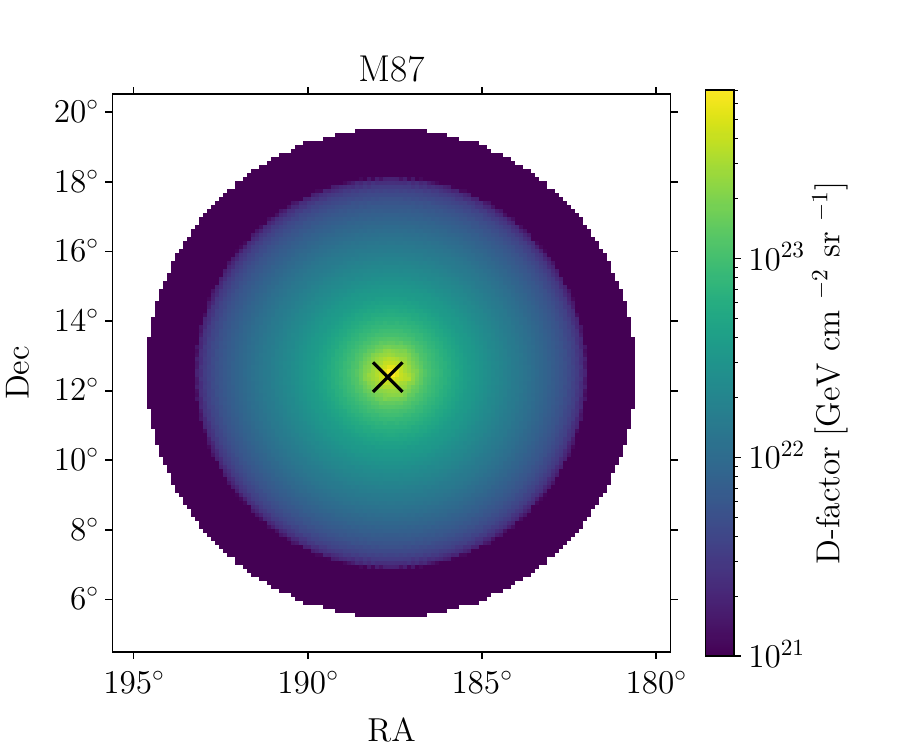}&
\includegraphics[width=0.47\textwidth]{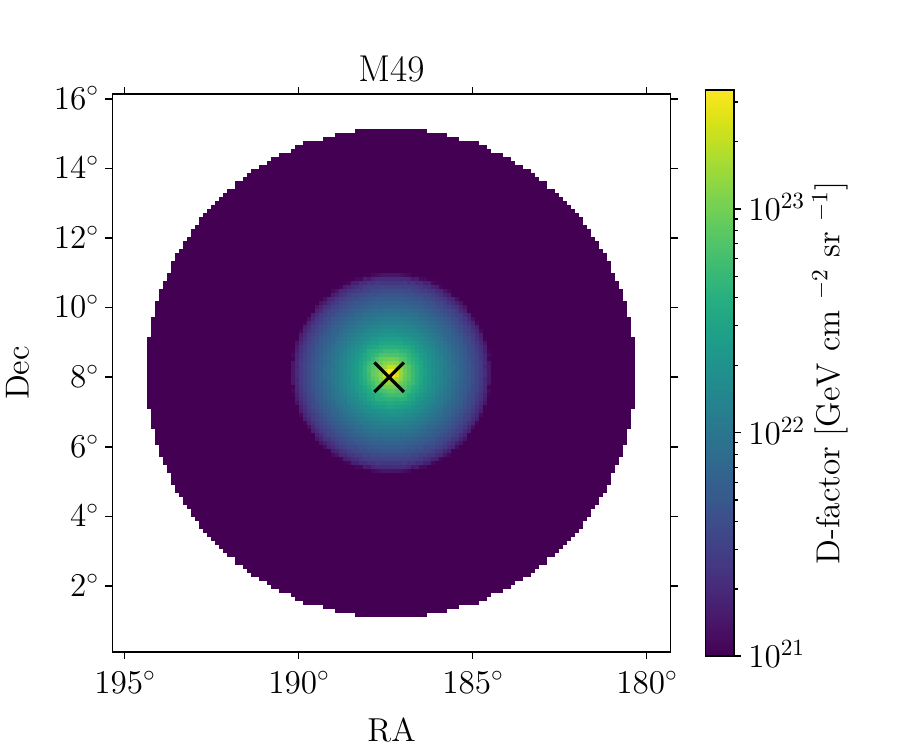}
\end{tabular}
}
\caption{The spatial templates or D-factors used in this work for the two sub-clusters. \textit{Left:} M87. \textit{Right:} M49.}
\label{fig:dfac}
\end{figure*}

\subsection{\label{subsec:dmmodel}Spatial and Spectral Model of Dark Matter}

\begin{table}
\centering
\begin{tabular}{ c | c | c | c | c | c}
\hline
 Object & Distance [Mpc] & $z$ & R$_{\rm vir}$ [kpc] & $\rho_s$ [M$_{\odot}$/kpc$^3$] & r$_s$ [kpc]\\
 \hline
 M87 & 17.2 & 0.00428 & 1700 & $6.96 \times 10^{5}$ & 403.8 \\
 M49 & 17.1 & 0.00327 & 880  & $1.41 \times 10^{6}$ & 157.7 \\
 \hline
 \end{tabular}
 \caption{
 The various properties of M87 and M49 used in the construction of D factors. Columns 2--6 list the distance, redshift, virial radius, scale density and scale radius respectively.
 }
\label{tab:tab1}
\end{table}
The flux of gamma rays from decaying DM in an astrophysical object is given by,

\begin{equation}
\label{eq:1}
\frac{d\phi}{dE} = \frac{1}{4\pi \tau M_{\chi}}\frac{dN}{dE}D \exp(-\tau_{\rm ebl}) ,
\end{equation}

where $\tau$ is the decay lifetime of DM, $M_{\chi}$ is the DM mass, $dN/dE$ is the gamma-ray spectrum per DM decay, and $D$ is known as the D-factor encoding the spatial distribution of DM in the target of interest. It is defined as the integral of the DM density $\rho_{\rm DM}$ along the line-of-sight (l.o.s.) and over the solid angle $\Delta\Omega $,
\begin{equation}
D = \int_{\Delta \Omega}  \int_{\rm l.o.s.} {\rm d}\Omega \, {\rm d} s \ \rho_{\rm DM}(s,\Omega).
\end{equation}

We consider decaying DM producing final-state photons via five different channels: $b \bar{b}$, $\tau \bar{\tau}$, $W^+W^-$, $\mu^+\mu^-$, and $t\bar{t}$. We consider DM masses between 1 TeV and 1 PeV. For each channel, we obtain the gamma-ray spectrum for decays from the publicly available HDM repository \cite{Bauer:2020jay}, incorporating the electroweak corrections. For each channel, we assume a 100\% branching ratio, i.e. the DM particle only decays to a given lepton, quark, or boson channel.

The D-factor at a given position in the region of interest depends on the assumed DM halo properties. The DM halo for galaxies consists of a smoothly distributed main halo as well as an additional substructure, that is attributed to the gravitationally clumped over-densities in the main halo. We construct a spatial template for the Virgo cluster as a combination of the DM templates for M87 and M49 using the software package CLUMPY \cite{Hutten:2018aix}. 

Each generated template
encompasses a region of interest with radius 7$^\circ$
and the combined M87-M49 template covers 10$^\circ$
in right ascension and 12$^\circ$
in declination. To generate these templates, we define the
parameters for the underlying DM distribution by referring to the main halo and substructure
properties inferred by the velocity profiles of the stars in the galaxies and N-body simulations \cite{Diemand:2006ik}. For the distribution of DM in the main halo and sub-haloes we use the generalized Navarro-Frenk-White (NFW) profile \cite{Navarro:1995iw}, with the values of free parameters fixed following Ref. \cite{Ackermann:2015fdi} (table \ref{tab:tab1}). Another important input is the subhalo concentration which is usually parametrized as a function of mass over-density within a fixed radius of the halo center \cite{Sanchez-Conde:2013yxa}, by fitting data from cosmological simulations. In this work, we adopt the characterization in Ref. \cite{Moline:2016pbm} that takes into account the spatial dependence of subhalos in field halos. Other characteristic properties of M87 and M49 used in the simulations are listed in table \ref{tab:tab1}.  The simulated D-factor distributions for the two subclusters are shown in figure \ref{fig:dfac}.  We note that compared to DM annihilation, the decay limits are relatively insensitive to underlying assumptions about the exact DM profile. The D-factor is primarlily determined by the mass of the underlying main halo as shown in the case of other extended objects analyzed with HAWC \cite{HAWC:2018eaa}.

The two D-factor templates for the two sub-clusters are added together and analyzed as a single extended source. The expected number of events in a given pixel comprising the DM template and the point source M87 is obtained by convolving the expected flux (eq. \ref{eq:1}) with the response of the HAWC detector at the given coordinates, for a fixed DM mass and decay lifetime. The expectation can then be compared to the null hypothesis which consists of events due to the isotropic background and any diffuse extragalactic emission.  
\begin{figure*}[t]
\makebox[0.8\width][c]{
\begin{tabular}{@{}cc@{}}
\includegraphics[width=0.47\textwidth]{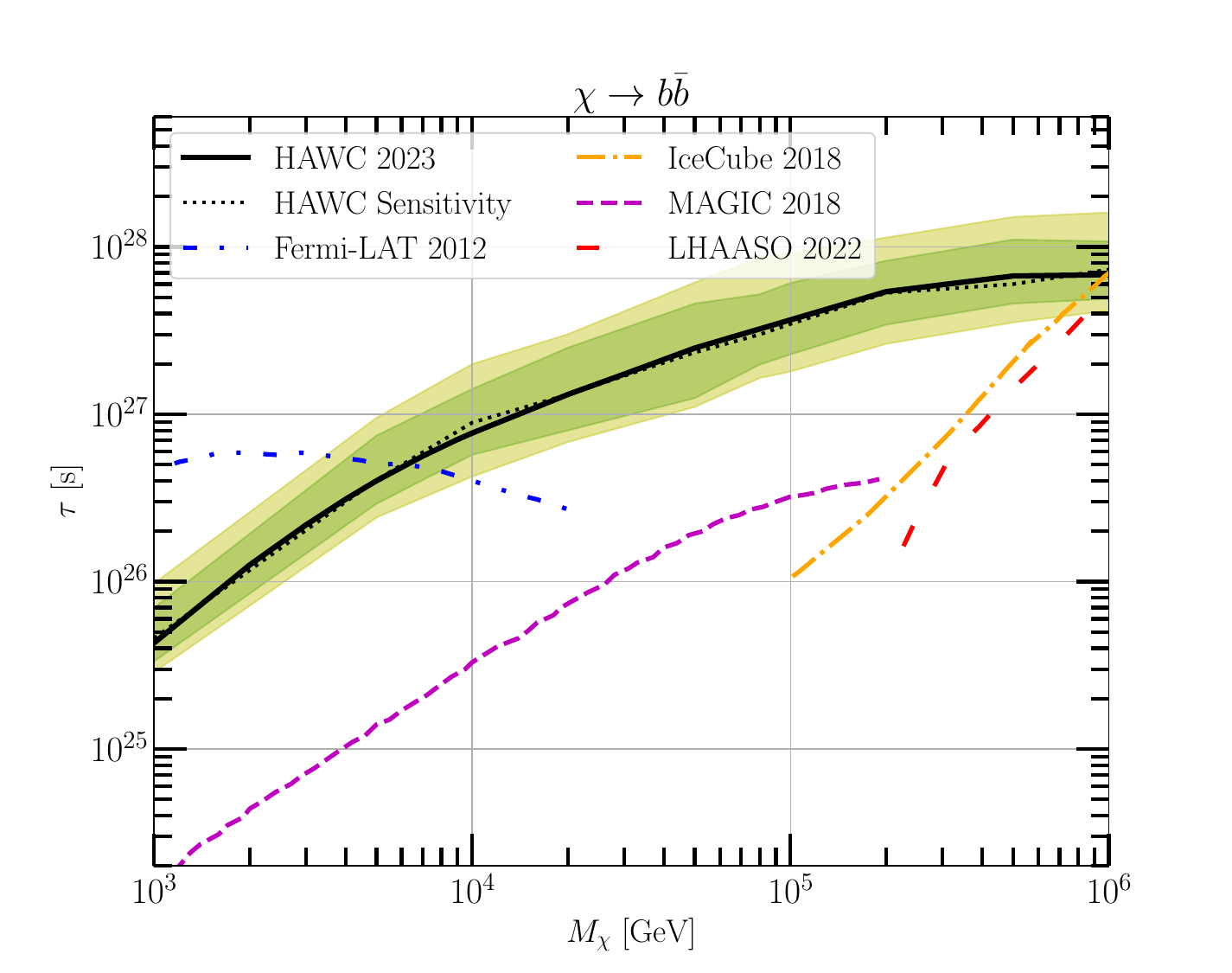} &
\includegraphics[width=0.47\textwidth]{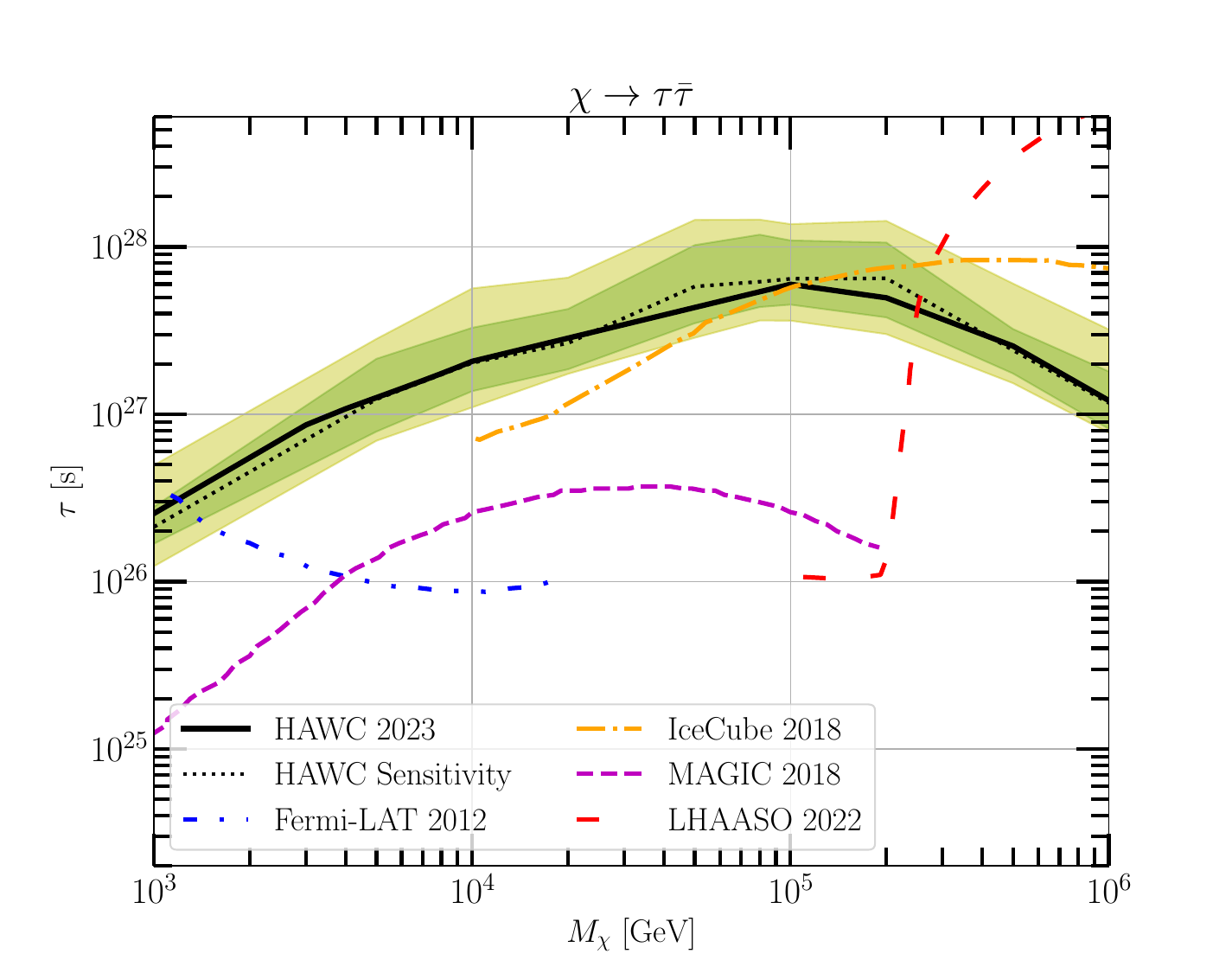} \\
\includegraphics[width=0.47\textwidth]{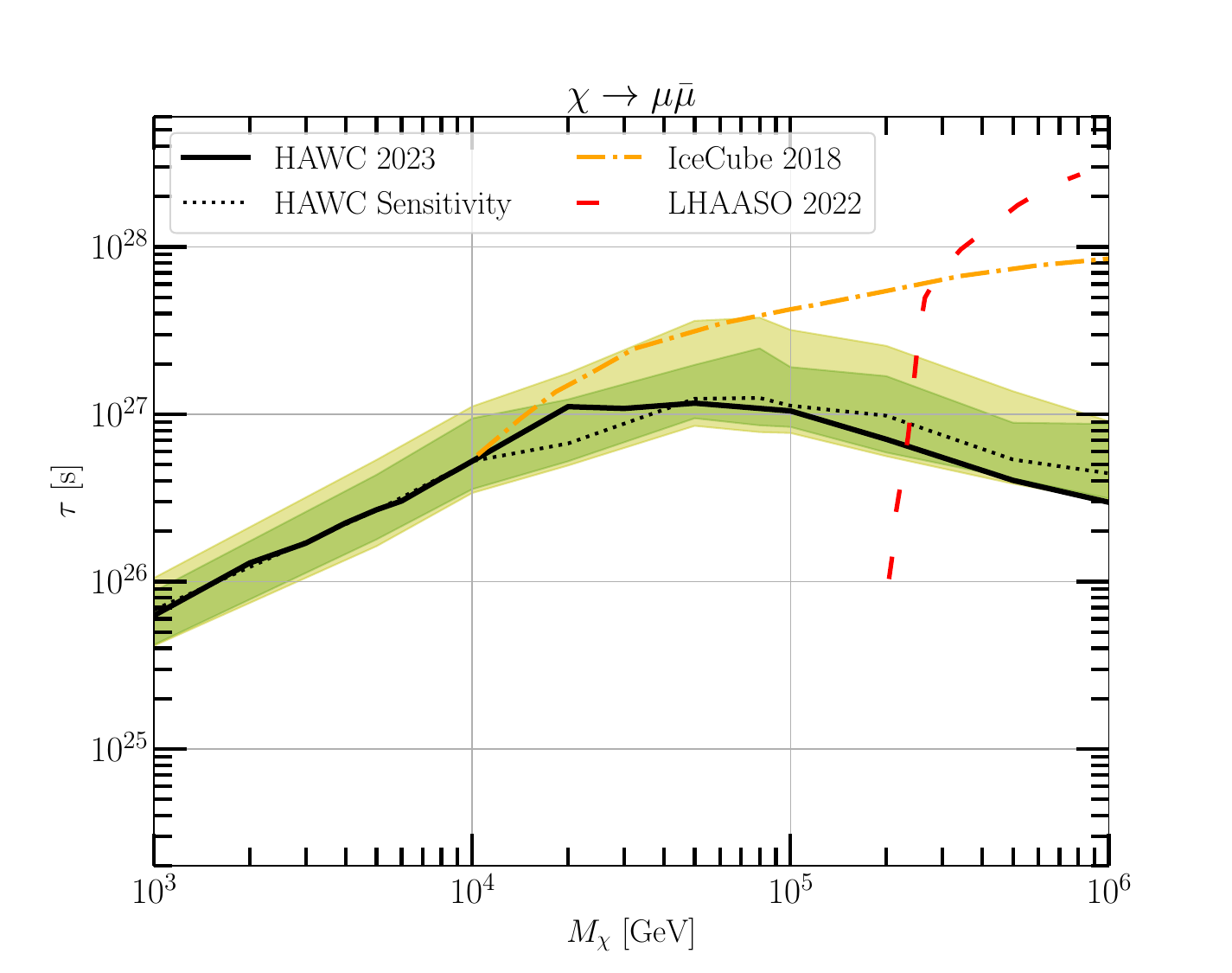} &
 \includegraphics[width=0.47\textwidth]{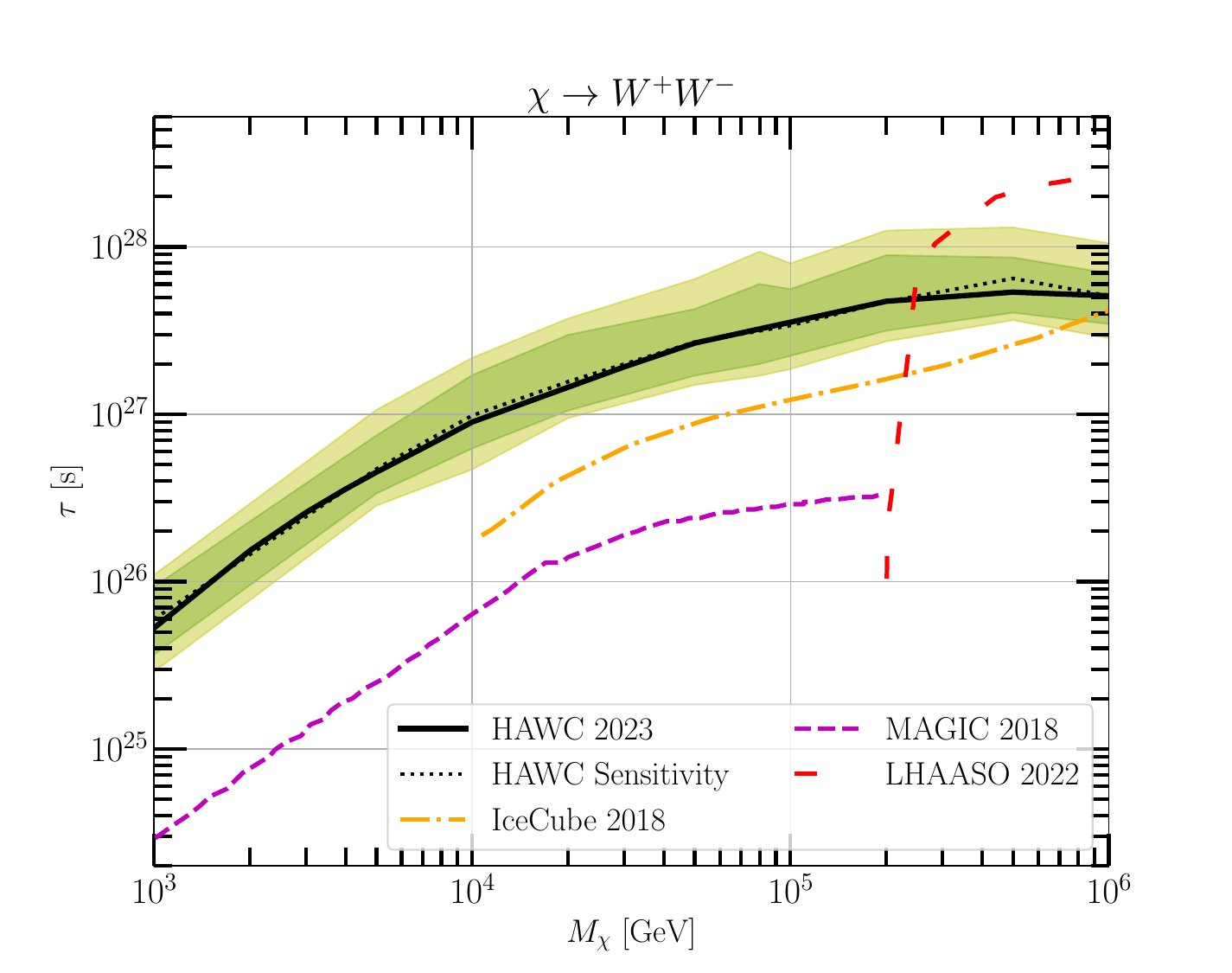}\\
\includegraphics[width=0.47\textwidth]{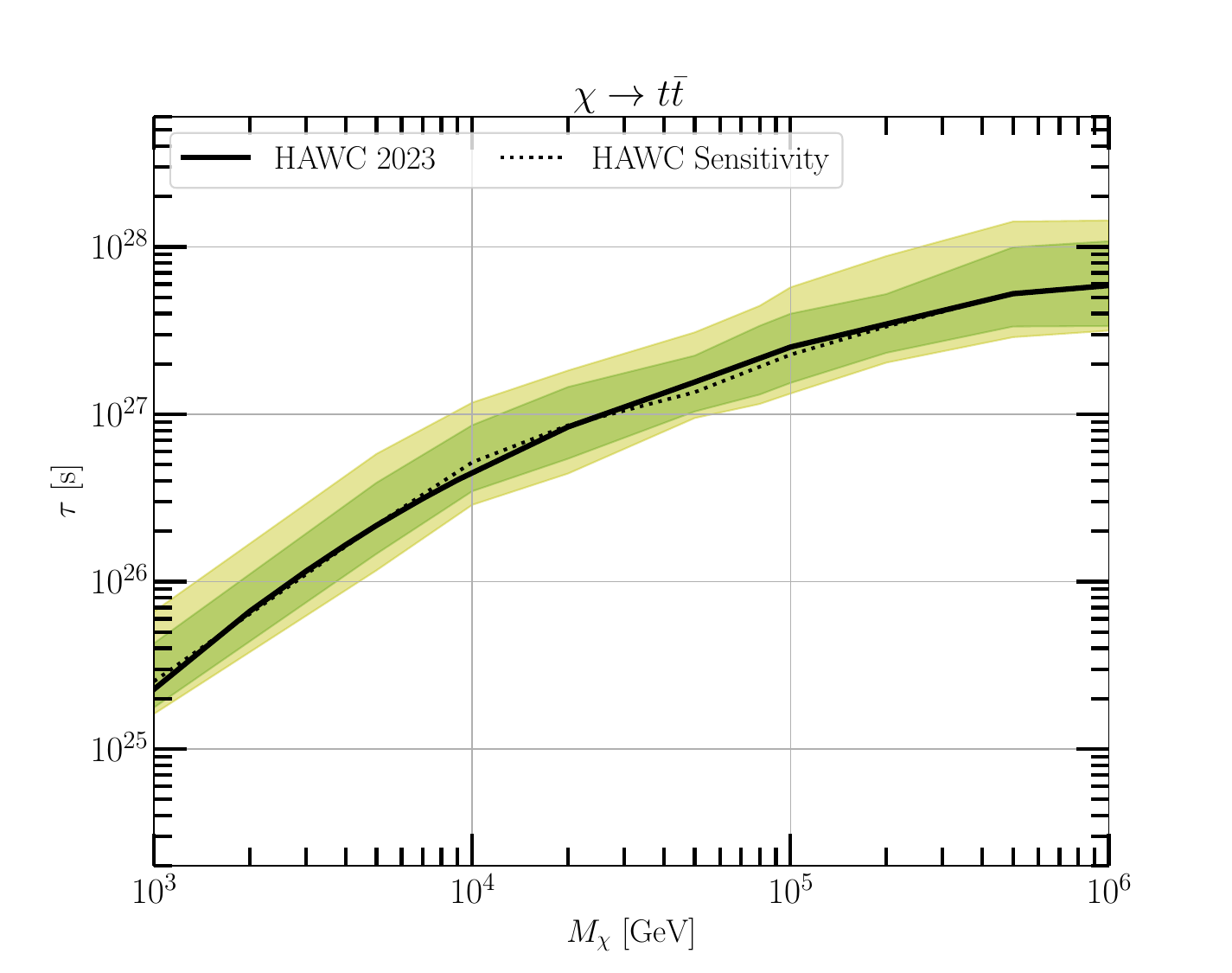} 
\end{tabular}
}
\caption{95\% confidence level lower limit constraints on the time of DM decay via $b\bar{b}$, $\tau\bar{\tau}$, $\mu^+\mu^-$, $W^+W^-$ and $t\bar{t}$. The HAWC results from the Virgo cluster and sensitivity (expected limits) are shown in the black solid and dotted lines respectively. The shaded bands indicate the central 68\% and 95\% expected limits. For comparison, results are also shown from IceCube \cite{IceCube:2018tkk}, Fermi-LAT \cite{2012ApJ...761...91A}, LHAASO \cite{LHAASO:2022yxw} and MAGIC \cite{MAGIC:2018tuz} collaborations.}
\label{fig:limits}
\end{figure*}

\section{\label{sec:res}Results and Conclusion }

No significant emission (beyond M87) is observed from the assumed DM morphology. The highest TS value for a fit to the DM hypothesis is 5.2 ($<2\sigma$) ,for 1 PeV dark matter decay to $\mu \bar{\mu}$. We, therefore, place lower limits at 95\% CL on the decay lifetime $\tau$ for every combination of DM mass and channel. Figure \ref{fig:limits} shows the resulting constraints on $\tau$ as a function of DM mass for all channels considered in this work. We also compute the sensitivity of the analysis by repeating the search on background-only regions in the same declination range as the Virgo cluster in 200 trials per mass-channel combination. In each trial, the background is poisson-fluctuated to produce a simulated dataset. Upper limits are obtained on the decay lifetime of DM following the method outlined above, and the median upper limit from these trials is considered to be the sensitivity of the analysis. The strongest constraints are obtained for DM decay to $b\bar{b}$ due to the soft spectrum of the channel, with much of the signal coming from multi-TeV energy photons within HAWC's sensitivity range. At the highest masses, the limits (and sensitivity) worsen.  Different EBL models differ in their prediction of the abosrption at a given energy. We note that the Franchescini model used in this work is a conservative choice \cite{Franceschini:2017iwq} when it comes to the effect on the constraints. Using a model such as in Ref. \cite{2010ApJ...712..238F} would improve the limits up to a factor $\sim 10$.   We also compare our results to limits obtained by IceCube \cite{IceCube:2018tkk}, Fermi-LAT \cite{2012ApJ...761...91A}, LHAASO \cite{LHAASO:2022yxw} and the MAGIC collaborations \cite{MAGIC:2018tuz} using nearby galaxies/clusters. As seen in the figures, HAWC limits are the strongest for the $W^+W^-$ channel for masses between 1 and 200 TeV. For $b\bar{b}$ and $\tau \bar{\tau}$, HAWC constraints are the strongest above $\sim$ 5 TeV to 100 TeV. HAWC continues to take data, and with the addition of outrigger tanks, will be able to extend its sensitivity to multi-PeV DM masses in the future. 

\clearpage
\begin{acknowledgements}
We acknowledge the support from: the US National Science Foundation (NSF); the US Department of Energy Office of High-Energy Physics; the Laboratory Directed Research and Development (LDRD) program of Los Alamos National Laboratory; Consejo Nacional de Ciencia y Tecnolog\'{i}a (CONACyT), M\'{e}xico, grants 271051, 232656, 260378, 179588, 254964, 258865, 243290, 132197, A1-S-46288, A1-S-22784, CF-2023-I-645, c\'{a}tedras 873, 1563, 341, 323, Red HAWC, M\'{e}xico; DGAPA-UNAM grants IG101323, IN111716-3, IN111419, IA102019, IN106521, IN110621, IN110521 , IN102223; VIEP-BUAP; PIFI 2012, 2013, PROFOCIE 2014, 2015; the University of Wisconsin Alumni Research Foundation; the Institute of Geophysics, Planetary Physics, and Signatures at Los Alamos National Laboratory; Polish Science Centre grant, DEC-2017/27/B/ST9/02272; Coordinaci\'{o}n de la Investigaci\'{o}n Cient\'{i}fica de la Universidad Michoacana; Royal Society - Newton Advanced Fellowship 180385; Generalitat Valenciana, grant CIDEGENT/2018/034; The Program Management Unit for Human Resources \& Institutional Development, Research and Innovation, NXPO (grant number B16F630069); Coordinaci\'{o}n General Acad\'{e}mica e Innovaci\'{o}n (CGAI-UdeG), PRODEP-SEP UDG-CA-499; Institute of Cosmic Ray Research (ICRR), University of Tokyo. H.F. acknowledges support by NASA under award number 80GSFC21M0002. We also acknowledge the significant contributions over many years of Stefan Westerhoff, Gaurang Yodh and Arnulfo Zepeda Dominguez, all deceased members of the HAWC collaboration. Thanks to Scott Delay, Luciano D\'{i}az and Eduardo Murrieta for technical support.
\end{acknowledgements}

\bibliography{mbib}

\end{document}